\documentstyle[12pt]{article}

\makeatletter
\if@twoside \oddsidemargin -17pt \evensidemargin 00pt
\else \oddsidemargin 00pt \evensidemargin 00pt
\fi
\newcount\@tempcntc
\def\@citex[#1]#2{\if@filesw\immediate\write\@auxout{\string\citation{#2}}\fi
  \@tempcnta\z@\@tempcntb\m@ne\def\@citea{}\@cite{\@for\@citeb:=#2\do
    {\@ifundefined
       {b@\@citeb}{\@citeo\@tempcntb\m@ne\@citea
        \def\@citea{,\penalty\@m\ }{\bf ?}\@warning
       {Citation `\@citeb' on page \thepage \space undefined}}%
    {\setbox\z@\hbox{\global\@tempcntc0\csname
b@\@citeb\endcsname\relax}%
     \ifnum\@tempcntc=\z@ \@citeo\@tempcntb\m@ne
       \@citea\def\@citea{,\penalty\@m}
       \hbox{\csname b@\@citeb\endcsname}%
     \else
      \advance\@tempcntb\@ne
      \ifnum\@tempcntb=\@tempcntc
      \else\advance\@tempcntb\m@ne\@citeo
      \@tempcnta\@tempcntc\@tempcntb\@tempcntc\fi\fi}}\@citeo}{#1}}

\def\@citeo{\ifnum\@tempcnta>\@tempcntb\else\@citea
  \def\@citea{,\penalty\@m}%
  \ifnum\@tempcnta=\@tempcntb\the\@tempcnta\else
   {\advance\@tempcnta\@ne\ifnum\@tempcnta=\@tempcntb \else
\def\@citea{--}\fi
    \advance\@tempcnta\m@ne\the\@tempcnta\@citea\the\@tempcntb}\fi\fi}

\makeatother

\expandafter\ifx\csname mathrm\endcsname\relax\def\mathrm#1{{\rm
#1}}\fi

\def\no{\nonumber\\}
\def\nl{\nonumber\\}
\def\nn{\nonumber}

\def\beq{\begin{equation}}
\def\eeq{\end{equation}}
\def\beqar{\begin{eqnarray}}
\def\eeqar{\end{eqnarray}}
\def\barr#1{\begin{array}{#1}}
\def\earr{\end{array}}

\def\bma{\begin{displaymath}}
\def\ema{\end{displaymath}}

\def\be{\beta}
\def\Ga{\Gamma}

\def\de{\delta}

\def\veps{\varepsilon}

\def\si{\sigma}
\def\Si{\Sigma}

\def\refeq#1{\mbox{Eq.~(\ref{#1})}}

\def\refeq#1{\mbox{(\ref{#1})}}

\def\refeqf#1{\mbox{(\ref{#1})}}
\def\refse#1{\mbox{Sect.~\ref{#1}}}
\def\citere#1{\mbox{Ref.~\cite{#1}}}
\def\citeres#1{\mbox{Refs.~\cite{#1}}}

\renewcommand{\L}{{\cal{L}}}

\def\mathswitchr#1{\relax\ifmmode{\mathrm{#1}}\else$\mathrm{#1}$\fi}
\newcommand{\Pf}{\mathswitch  f}
\newcommand{\PW}{\mathswitchr W}
\newcommand{\PZ}{\mathswitchr Z}

\newcommand{\PH}{\mathswitchr H}

\newcommand{\Pep}{\mathswitchr {e^+}}
\newcommand{\Pem}{\mathswitchr {e^-}}
\newcommand{\PWp}{\mathswitchr {W^+}}
\newcommand{\PWm}{\mathswitchr {W^-}}

\newcommand{\Ff}{\mathswitch f}
\newcommand{\Ffbar}{\mathswitch{\bar f}}
\newcommand{\FW}{\mathswitch W}
\newcommand{\FZ}{\mathswitch Z}
\newcommand{\FA}{\mathswitch A}

\def\mathswitch#1{\relax\ifmmode#1\else$#1$\fi}
\newcommand{\Mf}{\mathswitch {m_\Pf}}

\newcommand{\MW}{\mathswitch {M_\PW}}

\newcommand{\MZ}{\mathswitch {M_\PZ}}
\newcommand{\MH}{\mathswitch {M_\PH}}

\newcommand{\rw}{{\mathrm{W}}}
\newcommand{\sw}{\mathswitch {s_{\rw}}}
\newcommand{\cw}{\mathswitch {c_{\rw}}}

\newcommand{\Qf}{\mathswitch {Q_\Pf}}

\newcommand{\ri}{{\mathrm{i}}}
\renewcommand{\ri}{{i}}
\renewcommand{\d}{{\mathrm{d}}}

\newcommand{\ses}{self-energies}

\def\fbar{\bar f}
\def\Fhat{\hat F}
\def\Vhat{\hat V}
\def\What{\hat W}

\def\Zhat{\hat Z}
\def\Ahat{\hat A}

\def\Hhat{\hat H}
\def\Phihat{\hat \Phi}
\def\phihat{\hat \phi}
\def\vphihat{\hat \varphi}
\def\chihat{\hat \chi}
\def\thetahat{\hat \theta}
\def\SF{J_{\Fhat}}
\def\SFc{J_{\Fhat^\dagger}}
\def\Sf{J_f}
\def\Sfbar{J_{\fbar}}

\def\xiQ{\xi_Q}
\def\xiB{\xi_B}
\def\xiWhat{\hat\xi_W}
\def\xiAhat{\hat\xi_A}
\def\xiZhat{\hat\xi_Z}

\def\rY{{\mathrm{Y}}}
\def\rT{{\mathrm{T}}}

\newcommand{\bfm}{BFM}

\newcommand{\BF}{\mathrm{BF}}
\newcommand{\GF}{\mathrm{GF}}
\newcommand{\full}{{\mathrm{full}}}

\renewcommand{\Re}{\mathop{\mathrm{Re}}}
\newcommand{\Zc}{Z_{\mathrm{c}}}
\newcommand{\dgd}[1]{\frac{\de\Ga^\full}{\de#1}}

\def\draftdate{\relax}
\def\mda{\relax}
\def\mua{\relax}
\def\mla{\relax}
\def\draft{
\def\thtystars{******************************}
\def\sixtystars{\thtystars\thtystars}
\typeout{}
\typeout{\sixtystars**}
\typeout{* Draft mode!
         For final version remove \protect\draft\space in source file *}
\typeout{\sixtystars**}
\typeout{}
\def\draftdate{\today}
\def\mua{\marginpar[\boldmath\hfil$\uparrow$]%
                   {\boldmath$\uparrow$\hfil}%
                    \typeout{marginpar: $\uparrow$}\ignorespaces}
\def\mda{\marginpar[\boldmath\hfil$\downarrow$]%
                   {\boldmath$\downarrow$\hfil}%
                    \typeout{marginpar: $\downarrow$}\ignorespaces}
\def\mla{\marginpar[\boldmath\hfil$\rightarrow$]%
                   {\boldmath$\leftarrow $\hfil}%
                    \typeout{marginpar: $\leftrightarrow$}\ignorespaces}
\def\Mua{\marginpar[\boldmath\hfil$\Uparrow$]%
                   {\boldmath$\Uparrow$\hfil}%
                    \typeout{marginpar: $\Uparrow$}\ignorespaces}
\def\Mda{\marginpar[\boldmath\hfil$\Downarrow$]%
                   {\boldmath$\Downarrow$\hfil}%
                    \typeout{marginpar: $\Downarrow$}\ignorespaces}
\def\Mla{\marginpar[\boldmath\hfil$\Rightarrow$]%
                   {\boldmath$\Leftarrow $\hfil}%
                    \typeout{marginpar: $\Leftrightarrow$}\ignorespaces}
\overfullrule 5pt
\oddsidemargin -15mm
\marginparwidth 29mm
}

\makeatletter

\def\eqnarray{\stepcounter{equation}\let\@currentlabel=\theequation
\global\@eqnswtrue
\global\@eqcnt\z@\tabskip\@centering\let\\=\@eqncr
$$\halign to \displaywidth\bgroup\hskip\@centering
  $\displaystyle\tabskip\z@{##}$\@eqnsel&\global\@eqcnt\@ne
  \hskip 2\arraycolsep \hfil${##}$\hfil
  &\global\@eqcnt\tw@ \hskip 2\arraycolsep
$\displaystyle\tabskip\z@{##}$\hfil
   \tabskip\@centering&\llap{##}\tabskip\z@\cr}
\def\appendix{\par
 \setcounter{section}{0} \setcounter{subsection}{0}
 \def\thesection{\Alph{section}}}

\makeatother

\begin{document}
\thispagestyle{empty}
\def\thefootnote{\fnsymbol{footnote}}
\setcounter{footnote}{1}
\null
\hfill BI-TP 96/39 \\
\null
\hfill KA-TP-22-1996\\
\null
\hfill hep-ph/9609422\\
\null
\begin{center}
{\Large \bf The background-field formulation\\[.5em]
of the electroweak Standard Model}%
\footnote{Lecture given by G.\ Weiglein at the {\em XXXVI Cracow School
of Theoretical Physics}, Zakopane, Poland, June 1--10, 1996.}%
\footnote{Partially supported by the EC-network contract CHRX-CT94-0579.}
\vskip 2.5em
{\large 
{\sc Ansgar Denner}\\[1ex]
{\normalsize \it Paul-Scherrer-Institut, W\"urenlingen und Villigen,
Switzerland}\\[2ex]
{\sc Stefan Dittmaier}\\[1ex]
{\normalsize \it Theoretische Physik, Universit\"at Bielefeld,
Germany}\\[2ex]
{\sc Georg Weiglein}\\[1ex]
{\normalsize \it Institut f\"ur Theoretische Physik, Universit\"at
Karlsruhe, Germany}
}
\vskip 2em
\end{center} \par
\vskip 1.2cm
\vfil
{\bf Abstract} \par
The application of the background-field method to the
electroweak Standard Model and its virtues are reviewed.
Special emphasis is directed to the Ward identities that follow from
the gauge invariance of the background-field effective action. 
They are compatible with on-shell renormalization and imply a decent 
behavior of the background-field vertex functions.
Via the usual construction of connected Green functions
they transfer to Ward identities 
for connected Green functions which,
in distinction to the conventional formalism,
remain exactly valid in finite orders of perturbation theory even if a 
Dyson summation of self-energies (within a systematic use of
one-particle irreducible building blocks)
is performed.
Finally, we comment on the interplay between gauge invariance and
gauge-parameter \mbox{(in-)}dependence of vertex and Green functions
and the uniqueness of resummation procedures.
\par
\vskip 1cm
\null
\setcounter{page}{0}
\clearpage
\def\thefootnote{\arabic{footnote}}
\setcounter{footnote}{0}

\section{Introduction}

Gauge invariance is the guiding principle for the construction of 
the Standard Model of elementary particle physics.
However, in order to quantize gauge theories, 
gauge invariance 
is
broken by adding a gauge-fixing term to the Lagrangian.
The ambiguity in fixing the gauge introduces a gauge dependence 
of vertex and Green functions
which becomes manifest through the appearance of gauge parameters.
The gauge symmetry of the Lagrangian is restricted to BRS invariance 
which leads to complicated,
non-linear Slavnov--Taylor identities for Green functions. 

Physical observables, 
such as S-matrix elements, are 
gauge-independent 
in each complete order of perturbation theory.
However, 
the use of incomplete orders 
of the perturbative expansion is 
sometimes unavoidable. For example, the introduction of finite-width 
effects for unstable particles or of running couplings can only be 
achieved by a 
summation of self-energy corrections.
This so-called Dyson summation in general yields gauge-dependent answers 
in orders of perturbation theory that are not completely taken into account.
Moreover Dyson summation 
in general violates Slavnov--Taylor identities, and thus 
gauge cancellations and the consistency of the predictions may be destroyed.
 
\begin{sloppypar}
The situation is considerably improved in the 
framework of the
background-field method \cite{BFM,Ab81,Ab83} (BFM) where quantization is
performed without losing gauge invariance of the effective action. 
This manifests itself in simple (QED-like) Ward identities which imply a 
decent 
behavior of the vertex functions. Moreover, the Ward identities for
connected Green functions are not violated by 
a consistent Dyson summation.
In addition,
the BFM provides a 
number of technical advantages.
The purpose of this article is to review the basic features of the
formulation of the SM in the framework of the 
BFM \cite{BFMvPT,bgflong,bgfring,BFMWI}. 
\end{sloppypar}

Section~\ref{se:bfm} contains a discussion of the background-field
effective action and vertex functions. 
The construction of the gauge-invariant effective action is 
sketched
in \refse{se:Gamma}. The corresponding Ward identities for the vertex 
functions are considered in \refse{sec:WI}. These Ward identities
remain valid after
the usual on-shell renormalization if the field renormalization
is chosen appropriately, as described in \refse{se:ren}. As an 
illustration of the improved properties of BFM vertex functions,
in \refse{se:vfprop}
we define running couplings 
which possess a gauge-independent high-energy behavior and are governed by the
renormalization group.

Section~\ref{se:sme&gfs} deals with S-matrix elements and connected
Green functions. In \refse{sec:Smatr} we sketch the construction of
the generating functional for connected Green functions 
from which the S-matrix is obtained as usual by the reduction formula. 
This construction naturally introduces the full propagators (including the 
Dyson-summed self-energy corrections)
without violating the Ward identities for connected Green functions in 
finite orders of perturbation theory. This fact and some consequences are
explained in \refse{se:dyson}.

Section~\ref{se:BFMPT} provides a discussion of the connection between
the gauge-parameter \mbox{(in-)}\discretionary{}{}{}de\-pendence of 
vertex functions and 
the existence of 
Ward identities for these vertex functions
which are related to the invariance of the
classical Lagrangian.

\section{Effective action and vertex functions}
\label{se:bfm}
\subsection{The gauge-invariant effective action for the Standard Model}
\label{se:Gamma}

The BFM is a technique for
quantizing gauge theories without losing explicit gauge invariance
of the effective action~\cite{BFM,Ab81}.
This is done by decomposing the usual fields $\hat\varphi$ in the classical
Lagrangian $\L_{\mathrm{C}}$ into background fields $\hat\varphi$
and quantum fields $\varphi$,
\beq
\label{eq:quantLc}
{\cal L}_{\mathrm{C}}(\hat\varphi) \rightarrow 
{\cal L}_{\mathrm{C}}(\hat\varphi + \varphi).
\eeq
While the background fields are treated as external sources, only
the quantum fields are 
variables of integration in the functional integral. 
A gauge-fixing term is added which breaks only the
invariance with respect to quantum-field 
gauge transformations 
but retains the  invariance of the functional integral with respect to 
background-field gauge transformations. 
{}From the functional integral an effective action $\Ga[\hat\varphi]$ for
the background fields is
derived which is invariant under gauge transformations of the
background fields and thus gauge-invariant.

A detailed treatment of the 
SM, which 
has been presented in \citere{bgflong}, is 
beyond 
the scope of this short review.
Therefore, we restrict our discussion to the basic differences to
the conventional approach. While the gauge fields are treated as
specified in \refeq{eq:quantLc},
the complex scalar SU$(2)_{\rw}$ doublet field of the minimal
Higgs sector is written as the sum of a
background Higgs field $\hat\Phi$, 
having the usual non-vanishing
vacuum expectation value $v$, and a quantum Higgs field $\Phi$, 
whose vacuum expectation value is zero:
\begin{equation}
\label{eq:Phi}
\hat\Phi(x) = \left( \begin{array}{c}
\phihat ^{+}(x) \\ \frac{1}{\sqrt{2}}\bigl(v + {\hat H}(x) +i
\chihat (x) \bigr)
\end{array} \right) , \qquad
\Phi (x) = \left( \begin{array}{c}
\phi ^{+}(x) \\ \frac{1}{\sqrt{2}}\bigl(H(x) +i \chi (x) \bigr)
\end{array} \right) .
\label{eq:Hbq}
\end{equation}
Here ${\hat H}$ and $H$ denote the physical background and
quantum Higgs field, respectively, and $\phihat ^{+}, \chihat,
\phi ^{+}, \chi$ represent the unphysical Goldstone-boson fields.

The generalization of the 't~Hooft gauge fixing to the BFM~\cite{Sh81} reads
\beqar\label{tHgf}
\L_{\mathrm{GF}} &=& {}- \frac{1}{2\xiQ^W}
\biggl[(\de^{ac}\partial_\mu + g_2 \veps^{abc}\hat
W^b_\mu)W^{c,\mu}
       -ig_2\xiQ^W\frac{1}{2}(\hat\Phi^\dagger_i
\si^a_{ij}\Phi_j
                  - \Phi^\dagger_i
\si^a_{ij}\hat\Phi_j)\biggr]^2 \no
                 && {}- \frac{1}{2\xiQ^B}
\biggl[\partial_\mu B^{\mu}
       +ig_1\xiQ^B\frac{1}{2}(\hat\Phi^\dagger_i \Phi_i
                  - \Phi^\dagger_i \hat\Phi_i)\biggr]^2,
\eeqar
where 
$W_{\mu }^{a}$, $a$=1,2,3, represents the triplet of gauge 
fields associated with the weak isospin group SU$(2)_{\rw}$, and 
$B_{\mu }$ the gauge field  associated with the group
U$(1)_{\rY}$ of weak hypercharge $Y_{\rw}$.
The Pauli matrices are denoted by $\si^a$, $a=1,2,3$, 
and $\xiQ^W$ and
$\xiQ^B$ are parameters associated with the gauge fixing of the
quantum fields, one for SU$(2)_{\rw}$ and one for U$(1)_{\rY}$.
In order to avoid tree-level mixing between the quantum 
photon and $\PZ$-boson 
fields, we set $\xiQ=\xiQ^W=\xiQ^B$ in the following.
Background-field gauge invariance requires that the background gauge
fields appear only within a covariant derivative in the gauge-fixing term and
that the terms in brackets transform according to the adjoint
representation of the gauge group. The
gauge-fixing term 
of \refeq{tHgf} translates to the conventional one upon
replacing the background Higgs field by its vacuum expectation value and
omitting the background SU$(2)_{\rw}$ triplet field $\hat W^a_\mu$.

The vertex functions can be calculated from Feynman rules that
distinguish between quantum and
background fields. Whereas the quantum fields appear only inside
loops, the background fields are associated with external lines.
Apart from doubling of the gauge and Higgs fields, the
BFM Feynman rules differ from the conventional ones
only owing to the gauge-fixing and ghost terms. 
Because these terms are quadratic in the quantum fields, they affect 
only vertices that involve exactly two quantum fields and additional 
background fields.
Since the gauge-fixing term is non-linear in the fields, the
gauge parameter enters also the gauge-boson vertices.
The fermion fields can be treated as usual, i.e.\ 
they have the conventional Feynman rules,
and no distinction needs to be made between external and internal fields.
A complete set of BFM Feynman rules for the electroweak SM 
has been 
given in \citere{bgflong}.

The BFM was also applied to the non-linear scalar 
realization of the SM \cite{Di95}, which is physically equivalent to the 
linear scalar
representation \refeq{eq:Phi} but, e.g., more convenient for studying 
effects of a heavy Higgs-boson mass. In the following all formulae are 
given for the more familiar linear scalar realization.

\subsection{Ward identities} 
\label{sec:WI}

The invariance of the background-field effective action under
background-field gauge transformations with associated group parameters
$\thetahat^a$, 
\beq
\label{eq:WIGamma}
\frac{\delta\Gamma}{\delta\hat\theta^a} \;=\; 0, \qquad
a=A,Z,\pm,
\eeq
implies {\it linear} identities for the vertex functions that
are precisely the Ward identities related to the classical Lagrangian.
This is in contrast to the conventional formalism where, owing to
the gauge-fixing procedure, explicit gauge invariance is lost,
and Ward identities are obtained 
only from invariance under BRS transformations.
These Slavnov--Taylor identities have a more complicated 
non-linear
structure and in general involve ghost contributions%
\footnote{Note that also in the BFM a BRS invariance involving the
quantum fields is still valid and gives rise to Slavnov--Taylor
identities for Green functions with external quantum fields, which 
appear as substructures in the BFM vertex functions.}.

The BFM Ward identities are valid in all orders of perturbation theory
and hold for arbitrary values of the quantum gauge parameter $\xi_Q$.
We give some examples for illustration.
Concerning the notation and conventions for the vertex
functions we follow \citere{bgflong} throughout.
Some of the Ward identities for two-point functions involving
neutral gauge bosons read:
\beqar
\label{eq:sega}
k^\mu \Ga^{\Ahat\Ahat}_{\mu\nu}(k) = 0, \qquad
k^\mu \Ga^{\Ahat\Zhat}_{\mu\nu}(k) = 0, \qquad
k^\mu \Ga^{\Zhat\Zhat}_{\mu\nu}(k) -i\MZ \Ga^{\chihat\Zhat}_\nu(k) %= 0.
&=& 0, \nn\\ 
\label{eq:seZ2}
k^\mu \Ga^{\Zhat\chihat}_{\mu}(k) -i\MZ \Ga^{\chihat\chihat}(k)
+\frac{ie}{2\sw\cw} \Ga^{\Hhat}(0) &=& 0 . 
\eeqar
For the photon--fermion and the photon-\PW-boson vertices
QED-like Ward identities are derived, e.g.\
\beqar
\label{WIAff}
k^\mu \Ga^{\Ahat\Ffbar\Ff}_{\mu}(k,\bar p, p) &=& -e\Qf
\left[\Ga^{\Ffbar\Ff}(\bar p) - \Ga^{\Ffbar\Ff}(-p)\right],  \nn\\ 
\label{WIAWW}
k^\mu \Ga^{\Ahat\What^+\What^-}_{\mu\rho\si}(k,k_+,k_-) &=& 
e \left[\Ga^{\What^+\What^-}_{\rho\si}(k_+) -
\Ga^{\What^+\What^-}_{\rho\si }(-k_-)\right], \nn\\ 
k_+^\rho \Ga^{\hat\FA\hat\FW^+\hat\FW^-}_{\mu\rho\si}(k,k_+,k_-)
&-&\MW \Ga^{\hat\FA\hat\phi^+\hat\FW^-}_{\mu\si}(k,k_+,k_-) = 
\no && \hspace{-4em}
e\left[\Ga^{\hat\FW^+\hat\FW^-}_{\mu\si}(-k_-) -
\Ga^{\hat\FA\hat\FA}_{\mu\si}(k)
+\frac{\cw}{\sw} \Ga^{\hat\FA\hat\FZ}_{\mu\si}(k)\right] .
\label{WIWAW}
\hspace{2em}
\eeqar
Further Ward identities are listed in \citeres{bgflong,bgfring}.

\subsection{Gauge-invariant on-shell renormalization}
\label{se:ren}

In the on-shell renormalization scheme
the parameters $\MW$, $\MZ$, $\MH$, $\Mf$ are identified
with the physical masses (propagator poles),
and $e$ with the electric unit charge fixed in the Thomson limit. 
Of course this choice of physical parameters is still possible 
within the BFM.
However, the
BFM gauge invariance has important consequences for the structure of the
field renormalization constants necessary to render Green functions
and S-matrix elements finite.
The arguments which we give in the following are made
explicit for the one-loop level.%
\footnote{We implicitly assume the existence of an invariant
regularization scheme.}
It is easy, however, to extend them by induction to arbitrary
orders in perturbation theory. Because the renormalization of the 
fermionic sector is similar to the one
in the conventional formalism, we suppress it here.

We introduce the following renormalization constants for the parameters:
\beqar
{\MW^2}_{,0} &=& \MW^2 + \delta \MW^2 , \qquad
{\MZ^2}_{,0} = \MZ^2 + \delta \MZ^2 , \qquad
{\MH^2}_{,0} = \MH^2 + \delta \MH^2 , \no
e_0 &=& Z_e e = (1 + \delta Z_e) e , \qquad
t_0 = t + \delta t ,
\label{eq:renconsts1}
\eeqar
where the subscript ``0'' denotes bare quantities.
The tadpole counterterm $\delta t$ renormalizes the term
$t \Hhat(x)$
in the Lagrangian linear in the Higgs field $\Hhat$. 
It corrects for the shift in the minimum of the Higgs potential
owing to radiative corrections. Choosing $v$ as the correct vacuum
expectation value of the Higgs field $\Phihat$ is equivalent to
the vanishing of $t$.

Following the QCD treatment of~\citere{Ab81},
we introduce field renormalization only for the background fields,
\beqar
\What_{0}^{\pm}  & = & Z_{\What}^{1/2} \What^{\pm}
  = (1+\frac{1}{2}\delta Z_{\What}) \What^{\pm} , \no
\left(\barr{l} \Zhat_{0} \\ \Ahat_{0} \earr \right)  & = &
\left(\barr{ll} Z_{\Zhat\Zhat}^{1/2} & Z_{\Zhat\Ahat}^{1/2}  \\[1ex]
                Z_{\Ahat\Zhat}^{1/2} & Z_{\Ahat\Ahat}^{1/2}
      \earr
\right)
\left(\barr{l} \Zhat \\ \Ahat \earr \right)   =
\left(\barr{cc} 1 + \frac{1}{2}\delta Z_{\Zhat\Zhat} &
\frac{1}{2}\delta Z_{\Zhat\Ahat} \\ [1ex]
\frac{1}{2}\delta Z_{\Ahat\Zhat}  & 1 + \frac{1}{2}\delta
Z_{\Ahat\Ahat}
\earr \right)
\left(\barr{l} \Zhat \\[1ex] \Ahat \earr \right)  , \no
\hat S_{0} & = & Z_{\hat S}^{1/2} \hat S
 = (1+\frac{1}{2}\delta Z_{\hat S}) \hat S, \qquad
\hat S = \Hhat,\chihat,\phihat .
\label{eq:renconsts2}
\eeqar

In order to preserve the background-field gauge invariance,
the renormalized effective action has to be invariant under 
background-field gauge transformations.
This restricts the possible counterterms and relates the
renormalization constants introduced above. These relations can be
derived from the requirement that the
renormalized vertex functions fulfill Ward identities of the same form
as the unrenormalized ones.
As a consequence, also the counterterms have to fulfill these Ward
identities. An analysis of the Ward identities yields~\cite{bgflong}:
\beqar
\label{eq:delZB}
\delta Z_{\Ahat\Ahat} &=& - 2 \delta Z_e, \qquad
\delta Z_{\Zhat\Ahat} = 0, \qquad
\delta Z_{\Ahat\Zhat} = 2 \frac{\cw}{\sw}
    \frac{\delta \cw ^2}{\cw ^2} , \no
\delta Z_{\Zhat\Zhat} &=& - 2 \delta Z_e -
    \frac{\cw ^2 - \sw ^2}{\sw^2} \frac{\delta \cw ^2}{\cw ^2} , \qquad
\delta Z_{\What} = - 2 \delta Z_e -
    \frac{\cw ^2}{\sw^2} \frac{\delta \cw ^2}{\cw ^2} , \no
\delta Z_{\Hhat} &=& \delta Z_{\chihat} = \delta Z_{\phihat} %\no &=&
      = - 2 \delta Z_e -
	\frac{\cw ^2}{\sw^2} \frac{\delta \cw ^2}{\cw ^2} +
	\frac{\delta \MW^2}{\MW^2} ,
\eeqar
where
\beq
\cw^2 = \frac{\MW^2}{\MW^2} = 1-\sw^2, \qquad
\frac{\delta \cw ^2}{\cw ^2} =
\frac{\delta \MW^2}{\MW^2} - \frac{\delta \MZ^2}{\MZ^2} .
\eeq

The relations \refeq{eq:delZB} express the
field renormalization constants of all gauge bosons and scalars
completely in terms of the renormalization constants of the
electric charge and the particle masses. 
With this set of renormalization constants all background-field vertex 
functions become finite%
\footnote{Beyond one-loop
order one needs in addition a renormalization of
the quantum gauge parameters~\cite{Ab81}. At one-loop
level these counterterms
do not enter the background-field vertex functions, because
$\xi_Q$ does not appear in pure background-field vertices.
Clearly, the renormalization of gauge parameters is irrelevant
for gauge-independent quantities such as 
S-matrix elements at any order.}.
This is evident since the divergences of the vertex functions are
subject to the same 
restrictions as the counterterms.
In \citere{bgflong} it has been verified explicitly at one-loop
order that
a renormalization based on the on-shell definition of all
parameters can consistently be used in the BFM. It
renders all vertex functions finite while respecting the full
gauge symmetry of the BFM.

As the field renormalization constants are fixed by \refeq{eq:delZB},
the propagators in general acquire residues 
being different from unity but finite, 
and different fields can mix on shell. 
This is similar to the minimal on-shell
scheme of the conventional formalism~\cite{BHS}.
Therefore, when calculating S-matrix elements, one has to introduce 
(UV-finite) wave-function renormalization constants, which have been
explicitly given in \citere{BFMWI} for the gauge fields. 
However, just as in QED, the on-shell definition of the electric
charge together with gauge invariance automatically fixes the residue
of the photon propagator to unity. 

As a consequence of the relations between the renormalization constants,
the counterterm vertices of the background fields
have a much simpler structure than the ones in the conventional
formalism (see e.g.~\citere{Dehab}).
In fact, all vertices originating from a separately gauge-invariant term
in the Lagrangian acquire the same renormalization constants.
The explicit form of the counterterm vertices 
at one-loop order has been given in~\citere{bgflong}.

As the renormalized parameters are identified with the physical electron 
charge and the physical particle masses, they are manifestly
gauge-independent. Moreover, the original bare parameters in the
Lagrangian are obviously 
gauge-independent, as they represent free parameters of the
theory. The same is true for the bare charge
and the bare weak mixing angle as these are directly related to the free
bare parameters. Consequently, the counterterms
$\delta Z_e$ and $\delta\cw^2$ for the gauge couplings are
gauge-independent. The relations 
\refeq{eq:delZB} therefore imply that the field renormalizations
of all gauge-boson fields are gauge-independent. This is in
contrast to the conventional formalism where the field
renormalizations in the on-shell scheme are gauge-dependent.%
\footnote{In contrast to
$\delta Z_e$ and $\delta\cw^2$ the mass counterterms 
are not gauge-independent. The bare masses depend on the bare
vacuum expectation value $v_0$ of the Higgs field,
which is not a free parameter of the theory. See \citere{bgflong} for a
discussion.}

\subsection{Properties of background-field vertex functions}
\label{se:vfprop}

The QED-like Ward identities valid for the BFM vertex 
functions (for all
values of $\xiQ$) give rise to improved theoretical properties of
form factors defined within the BFM compared to their conventional
counterparts~\cite{BFMvPT,bgflong}. 

As an example, we consider the asymptotic behavior of gauge-boson
self-energies. Just as in QED, one can define running couplings in the 
BFM for the SM via na{\"\i}ve Dyson summation of \ses\ as follows:
\beqar
\label{eq:runcoupl}
e^2(q^2) &=& \frac{e_0^2}{1 + \Re \Pi^{\Ahat\Ahat}_0(q^2)}
         = \frac{e^2}{1 + \Re \Pi^{\Ahat\Ahat}(q^2)} , \no
g_2^2(q^2) &=& \frac{g_{2,0}^2}{1 + \Re\Pi^{\What\What}_0(q^2)}
         = \frac{g_2^2}{1 + \Re\Pi^{\What\What}(q^2)} ,
\eeqar
where 
$g_{2,0}=e_0/s_{\rw,0}$ and $g_2=e/\sw$.
The quantities $\Pi^{\Vhat\Vhat'}$ are related to the transverse parts of
the gauge-boson \ses\ 
$\Sigma^{\Vhat\Vhat'}_\rT$ as follows:
\beq
\Pi^{\Vhat\Vhat'}(q^2) =
\frac{\Sigma^{\Vhat\Vhat'}_\rT(q^2)-\Sigma^{\Vhat\Vhat'}_\rT(0)}{q^2}.
\eeq
The relations \refeq{eq:delZB} give rise to 
the following
properties of these running couplings: As indicated
in \refeq{eq:runcoupl}, the renormalization constants
cancel. Consequently, the running couplings are finite without renormalization
and thus independent of the renormalization scheme (as long as it respects
BFM gauge invariance). Their asymptotic behavior is gauge-independent and
governed by the renormalization group. In particular, the
coefficients of the leading logarithms in the \ses\ are equal to the
ones appearing in the $\be$-functions associated with the running
couplings. All these properties are completely analogous to those of the
running coupling in QED; they follow in the same way from the relations
\refeq{eq:delZB} as in QED from 
$Z_e = Z_{\Ahat\Ahat}^{-1/2}$. % and $Z_{g_2} = Z_{\What}^{-1/2}$,

\section{S-matrix and connected Green functions}
\label{se:sme&gfs}

\subsection{Construction}
\label{sec:Smatr}

S-matrix elements and connected Green functions are constructed by 
forming trees with vertex functions 
from the effective action $\Gamma[\hat\varphi]$ 
joined by background-field propagators. These
propagators are defined by adding a 
gauge-fixing term to $\Gamma[\hat\varphi]$, resulting in
\beq \label{eq:Gammafull}
\Ga^\full = \Ga + \ri\int\d^4x\, \L^{\BF}_{\GF}.
\eeq
The gauge-fixing term $\L^{\BF}_{\GF}$
is not related to the term \refeq{tHgf} that fixes the gauge 
of the quantum fields,
and the associated gauge parameters $\xiB^i$ 
enter only tree-level quantities but not the higher-order contributions 
to the vertex functions.

The generating functional of connected Green functions, $\Zc$, is
obtained from $\Ga^\full$ 
(as usual) by a Legendre transformation \cite{BFMWI},
\beq\label{eq:Zc}
\Zc[\SF,\Sf,\Sfbar] = \Ga^\full[\Fhat,f,\fbar]
+ \ri \int\d^4x \Bigl[\sum_{\Fhat} \SFc\Fhat
+ \sum_f (\fbar \Sf  + \Sfbar f) \Bigr]
\hspace{2em}
\eeq
with
$\hat F=\Ahat,\Zhat,\What^+,\What^-,\Hhat,\chihat,\phihat^+,\phihat^-$,
where $\hat F^\dagger$ denotes the complex conjugate of $\hat F$, and
\beqar\label{eq:GtoJ}
\ri \SFc = -\dgd{\hat F}, \qquad
\ri J_{\fbar} = \dgd{f}, \qquad
\ri J_f = -\dgd{\fbar}.
\eeqar

As a consequence,
the 1-particle reducible Green functions and S-matrix elements are
composed
as in the conventional formalism from a tree structure of vertex
functions. While the vertices in these trees are directly given by
the background-field vertex functions, the propagators are determined
as the inverse of the two-point vertex functions resulting from
$\Ga^{\full}$.
Note that these propagators contain by construction all self-energy
insertions. 
The Dyson summation of self-energy corrections has already been taken care of 
by this formalism. Of course, one can expand
the propagators and recover the ordinary perturbative expansion.

The S matrix follows from $\Zc$ by the usual reduction formula.
The equivalence of the S-matrix in the \bfm\
to the conventional one has been proven in 
\citeres{Ab83,Re85}.

Despite the distinction between background
and quantum fields, calculations in the BFM become in general simpler
than in the conventional formalism.
This is in
particular the case in the 't~Hooft--Feynman gauge ($\xiQ=1$)
for the quantum fields where many vertices simplify.
Moreover, the gauge fixing of the background fields is totally unrelated
to the gauge fixing of the quantum fields~\cite{Re85}. 
This freedom can be
used to choose a particularly suitable background gauge,
e.g.~the unitary gauge. % or a non-linear gauge~\cite{Ga81}.
In this way the number of Feynman diagrams can  considerably be reduced.

The Ward identities \refeqf{eq:WIGamma} for the effective action $\Ga$,
which generates the vertex functions,
translate into Ward identities for the functional $\Zc$, which generates
the connected Green functions. These Ward identities
were explicitly derived in \citere{BFMWI}
in a 't Hooft gauge for the background fields.
The (renormalized) two-point functions involving neutral 
gauge bosons obey for instance:
\beqar
k^\mu G^{\Ahat\Ahat}_{\mu\nu}(k) =  \frac{-\ri\xiAhat k_\nu}{k^2}, \qquad
 k^\mu G^{\Ahat\Zhat}_{\mu\nu}(k) = 0, \quad&\ &
k^\mu G^{\Zhat\Zhat}_{\mu\nu}(k) +i\xiZhat\MZ G^{\chihat\Zhat}_\nu(k)
= \frac{-\xiZhat\ri k_\nu}{k^2-\xiZhat\MZ^2}, \nl
k^\mu G^{\Zhat\chihat}_{\mu}(k) +i\xiZhat\MZ G^{\chihat\chihat}(k)
&=& \frac{-\xiZhat\MZ}{k^2-\xiZhat\MZ^2}.
\eeqar
For the photon--fermion and the photon-\PW-boson vertices we find
\beqar
\frac{\ri}{\xiAhat}k^2 k^\mu G^{\Ahat\Ffbar\Ff}_{\mu}(k,\bar p, p) &=& -e\Qf
\left[G^{\Ffbar\Ff}(\bar p) - G^{\Ffbar\Ff}(-p)\right],  \nn\\ 
\frac{\ri}{\xiAhat}k^2 k^\mu G^{\Ahat\What^+\What^-}_{\mu\rho\si}(k,k_+,k_-)&=& 
e \left[G^{\What^+\What^-}_{\rho\si}(k_+) -
G^{\What^+\What^-}_{\rho\si }(-k_-)\right] \nn\\ 
&& \hspace{-7em}
{}+e\frac{1}{k_+^2-\xiWhat\MW^2}k_{+,\rho}
\left[k_+^\mu G^{\What^+\What^-}_{\mu\si}(-k_-)
      +\xiWhat\MW  G^{\phihat^+\What^-}_{\si}(-k_-) \right]
\nn\\ && \hspace{-7em}
{}-e\frac{1}{k_-^2-\xiWhat\MW^2}k_{-,\si}
\left[k_-^\mu G^{\What^+\What^-}_{\rho\mu}(k_+)
      -\xiWhat\MW  G^{\What^+\phihat^-}_{\rho}(k_+) \right],
\eeqar
where we have used a Ward identity for the \PW-boson two-point function to 
simplify the last equation. The terms in the last line result from the 
gauge-fixing of the background fields.

After amputating the Green functions and putting fields on shell many
terms drop out in the Ward identities. Denoting amputated Green functions 
by $G_{\hat\phi_i\hat\phi_j\ldots}$,
we find for example the following identities
\beq\label{eq:WIET}
k^\nu G_{\Ahat\ldots,\nu} = 0, \qquad
k^\nu G_{\Zhat\ldots,\nu} = \ri\MZ G_{\chihat\ldots}, \qquad
k^\nu G_{\What^\pm\ldots,\nu} = \pm\MW G_{\phihat^\pm\ldots},
\eeq
where the ellipses stand for any on-shell fields. The first of
these identities expresses electromagnetic current conservation, the
others imply the well-known Goldstone-boson equivalence theorem,
as discussed in \citere{BFMWI} in detail.

\subsection{Dyson summation without violating Ward identities}
\label{se:dyson}

A particularly important property of the BFM is the fact that the 
BFM Ward identities for connected Green functions are not violated 
even in finite orders of perturbation theory by Dyson summation of 
self-energies, as was proven in \citere{BFMWI}. 
This is in contrast to the Slavnov--Taylor identities, which in general
only hold for connected Green functions in a given order of perturbation
theory if all contributions, including the propagators, are expanded
up to this order.

The crucial difference with respect to the Slavnov--Taylor identities
lies in the fact that the BFM Ward identities for vertex
functions $\Ga^{\vphihat_i\vphihat_j\cdots}$ are {\it linear} in all vertex
functions. Consequently they are exactly valid loop order by loop order
and the background-field effective action truncated at $n$-loop order,
$\Ga|_{n\mathrm{-loop}}$, is exactly gauge-invariant. Thus, the connected Green 
functions defined from 
$\Ga|_{n\mathrm{-loop}}$ via a Legendre transformation 
fulfill exactly the same Ward identities as those defined from
the full effective action containing all orders.
This implies that the Ward identities valid for the full connected 
Green functions in the BFM 
also hold exactly
for any fixed loop order in perturbation theory 
if the inverse propagators, which are just the two-point vertex functions,
are calculated in the same loop order as all other vertex functions.
This means that in the BFM Dyson summation does not destroy the
Ward identities for connected Green functions and the related gauge
cancellations at high energies. As an immediate consequence, the
Goldstone-boson equivalence theorem is valid in the BFM 
after Dyson summation.

Dyson summation is of particular importance for the treatment
of finite-width effects of unstable particles. The finite width of a
particle $P$ is introduced in field theory by Dyson summing the
self-energy $\Si^{PP}(k^2)$,
\beqar
-\left[\Ga^{PP}(k^2)\right]^{-1} &=& 
\frac{i}{k^2-M^2} + \frac{i}{k^2-M^2}i\Si^{PP}(k^2)\frac{i}{k^2-M^2} +
\cdots \nn\\
&=& \frac{i}{k^2-M^2+\Si^{PP}(k^2)},
\eeqar
and relating the finite width to the imaginary part of the self-energy.
However, since the summation mixes different orders in perturbation
theory, the result in general will not be gauge-invariant
in finite orders of perturbation theory.

This problem has been investigated recently in connection with 
the process $\Pep\Pem\to\PWp\PWm\to4f$, where a finite 
width has to be introduced for the \PW~boson.
In lowest order the W~boson decays only into fermions,
i.e.\ only fermion loops contribute to the relevant imaginary part
of the one-loop W-boson self-energy. The same 
is true for the \PZ~boson.
In \citere{FLscheme} it was argued that finite-width effects
of W and Z~bosons
can be introduced in tree-level amplitudes 
without destroying the Ward identities 
(and thus also the gauge cancellations)
by Dyson-summing the fermion-loop contributions to the 
self-energies {\em and\/} including also all the other fermion-loop
contributions in one-loop order%
\footnote{In the process $\Pep\Pem\to\PWp\PWm\to4f$ this amounts to 
inclusion of the fermion-loop corrections to the triple-gauge-boson vertex.}.

Within the framework of the BFM it is easy to understand why this 
prescription indeed preserves the Ward identities. 
The fermion-loop contributions at one-loop order 
in the conventional formalism coincide with those in the BFM, and 
as explained above in the BFM Dyson summation does not violate the Ward 
identities for connected Green functions. 

Obviously, the same procedure could be used within
the BFM for the general case, where also bosonic loop corrections 
contribute to the imaginary part of the self-energy. 
In contrast to the conventional formalism Dyson summing
the complete fermionic 
and bosonic corrections
in the BFM still preserves the Ward identities for connected Green functions.
However, both in the BFM and in the conventional formalism
one is in general faced with a gauge-parameter dependence at 
the incompletely calculated loop level.
As discussed in the following section, so
far---to the best of our knowledge---no
prescription is available that yields a unique unambiguous result
in the general case.

\section{Gauge invariance versus gauge-parameter independence}
\label{se:BFMPT}

Equipped with the gauge-invariant BFM effective action it is interesting
to investigate the connection between gauge invariance and
gauge-parameter \mbox{(in-)}dependence of vertex functions.
Motivated by the gauge independence of complete S-matrix elements,
several authors have performed
rearrangements of gauge-dependent parts between different vertex
functions resulting in definitions of separately
gauge-parameter-independent building 
blocks~\cite{Ke89,pinch}.
In particular,
the so-called pinch technique (PT)~\cite{pinch} 
provides
a quite general algorithm
for a rearrangement at the one-loop level which leads to 
gauge-parameter-free ``vertex functions'' with improved theoretical 
properties.
However, 
having no solid field-theoretical basis,
the PT suffers
from a number of conceptual and technical problems, like the unclear
field-theoretical meaning of building blocks constructed by rearranging
parts between different vertex functions. 
Moreover,
 the question of universality
and process-independence of the so-defined quantities could only be
verified by additional assumptions or a (necessarily incomplete)
case-by-case study.
Finally, the generalization of these methods to higher orders is not
straightforward.

It was shown in \citere{BFMvPT} that the PT ``vertex functions'' 
coincide with the special case $\xiQ = 1$ of the corresponding 
BFM vertex functions%
\footnote{In QCD this fact was also pointed out in \citere{Ha94}.}
and that the improved theoretical properties of the constructed building
blocks are a consequence of simple classical Ward identities.
In the BFM, these Ward identities are a direct
consequence of the gauge invariance and  hold in all orders of
perturbation theory.
It is instructive to investigate the origin of these Ward identities
within the PT. The crucial observation is that in this formalism the 
S-matrix elements are composed of 
gauge-parameter-free
``vertex functions'' connected by 
gauge-parameter-dependent 
tree-level propagators.
As the complete S-matrix element is independent of the
gauge parameters, certain non-trivial symmetry
relations between the new ``vertex
functions'' must exist that enforce the cancellation of the
remaining gauge-parameter dependence. This fact together with some
additional assumptions
on the independence of propagator-, vertex-, and box-like structures, 
as explained in some detail in \citere{bgfring},
leads to PT ``vertex functions'' that fulfill the classical Ward identities.
Note that the Ward identities do not uniquely fix these ``vertex functions'',
since one can always shift appropriate parts between the ``vertex 
functions'' that by themselves fulfill the Ward identities.

Thus, the validity of these non-trivial symmetry relations is not based on
the actual gauge-parameter independence of the new ``vertex functions'',
but---more generally---on the independence of the gauge 
parameters in the tree-level propagators from the gauge fixing
within loop diagrams. The prescription given within the PT is
just a special case of decoupling the gauge fixing in the loops from the
tree lines, like it is the case in the BFM. 
{}From these considerations it should be clear that, 
as far as gauge invariance and gauge or prescription independence is
concerned, application of methods like the PT {\em
within\/} the BFM is not meaningful, since the gauge fixings in the
loops and tree lines are already decoupled, and 
the elimination of $\xiQ$ 
can {\em not} be distinguished from
trivially putting $\xiQ$ to any specific value. 
In this context it is interesting to note
that a generalized PT algorithm was proposed in \citere{Pi96} which
reproduces the BFM vertex functions of QCD for arbitrary quantum
gauge parameter $\xi_Q$ at one loop.
This shows explicitly how the gauge dependence of the BFM vertex functions 
corresponds to an arbitrariness in fixing the PT algorithm.

As explained above, once resummations are involved, physical predictions 
in fixed orders of perturbation theory depend on the
gauge and any other prescription used to define the vertex functions.
This raises the question whether one of these prescriptions
is distinguished on physical grounds.
Because the BFM Ward identities, and thus the decent theoretical 
properties of the BFM (or PT) vertex functions, hold equally
well for any choice of $\xiQ$,
these Ward identities are not sufficient to provide a distinction.
The authors of \citere{Pa96} argue that the PT (or equivalently the BFM
with $\xi_Q=1$) is distinguished. 
Their only argument for rejecting the BFM for $\xi_Q\ne1$ is
the appearance of unphysical thresholds in the corresponding 
vertex functions. For $\xi_Q=1$ the unphysical thresholds happen
to appear at the same locations
as the physical thresholds and cannot be
distinguished in Green functions.
The starting point of the PT are S-matrix elements which evidently
do not involve unphysical thresholds. However, the  PT ``vertex
functions'' result from a split of the S-matrix elements into
propagator-, vertex- and box-like contributions.
It is by far not obvious that this split does not introduce unphysical
thresholds in the individual contributions,
which appear at the same locations as the physical thresholds.

As long as no 
physically distinguished
definition of vertex functions can be found, 
the existence of various
prescriptions signals the inherent ambiguity in defining form factors, 
resummations etc.\ on the basis of off-shell vertex functions. 
In view of applications for resummations of
bosonic loop contributions, for instance, this means that the ambiguity
found there is
not removed by a prescription like the PT but is only
traded on cost of the specific definition used to eliminate the gauge
parameters.
Finally we note that 
in a different context the authors of \citere{Je86} also arrived at the
conclusion that off-shell quantities are ambiguous even if gauge
invariance is imposed.

\section{Conclusion}
\label{concl}

We have reviewed some basic features of the application of the
background-field method (BFM) to the electroweak Standard Model (SM).

The gauge invariance of the BFM effective action implies simple
(QED-like) Ward identities for the vertex functions, which 
as a consequence possess
desirable theoretical properties like an improved high-energy, UV  and
IR behavior.
The BFM gauge invariance not only admits the usual on-shell 
renormalization but even simplifies its technical 
realization.
Moreover, the formalism provides additional advantages such as
simplifications in 
the Feynman rules and 
the possibility to use different gauges for tree and loop lines in
Feynman diagrams,
 thus allowing to  reduce the number of graphs. 

In contrast to the 
Slavnov--Taylor identities, the BFM
Ward identities 
are not violated by Dyson summation
if the connected Green functions are 
constructed from 
the complete set of 
vertex functions of a fixed loop order.
Consequently, gauge cancellations, and in particular the Goldstone-boson
equivalence theorem, are not disturbed if 
Dyson summation is applied.
This fact is important 
for the incorporation of 
finite-width effects of unstable particles 
within perturbation theory which requires a summation of self-energy 
corrections. Despite of this
important improvement in comparison to the conventional formalism, also
in the BFM a gauge-parameter dependence remains at the 
incompletely calculated loop level.
At present 
it is not known how or whether at all this problem can be avoided.

The decoupling of the different gauge fixings of tree and loop lines
does not uniquely determine the BFM vertex functions, as
already 
signaled by their dependence on the quantum gauge parameter
$\xi_Q$. This kind of ambiguity is 
also inherent in all 
those 
methods that eliminate the gauge-parameter dependence from vertex 
functions by redistributing the gauge-dependent parts. This is
in particular the case for the pinch-technique algorithm which 
reproduces the choice $\xi_Q=1$ of the BFM. Indeed 
the improved behavior of the 
BFM or pinch-technique
``vertex functions'' can be traced back to the 
Ward identities,
which hold in the BFM for arbitrary $\xi_Q$.
These Ward identities
follow in the BFM directly from gauge invariance,
but can only be derived in the pinch technique on the basis of
additional assumptions.

In conclusion, the BFM provides an alternative framework for quantizing 
gauge theories which compared to the conventional method has several
advantages both on conceptual and technical grounds.

\section*{Acknowledgements}

G.\ W.\ thanks M.\ Jezabek and the organizers of the school for the kind
invitation, the excellent organization and their hospitality during the
school.

\end{document}